\documentclass[letterpaper]{article}

\usepackage[T1]{fontenc}

\usepackage{geometry}
\usepackage{setspace}

\usepackage[articletitle = true,  maxauthors = 0, doi = true]{achemso}

\usepackage{graphicx}
\usepackage{float}
\newfloat{scheme}{htbp}{los}
\floatname{scheme}{Scheme}
\floatname{chart}{Chart}
\newfloat{graph}{htbp}{loh}

\usepackage{chemformula} 
\usepackage[version = 4]{mhchem} 

\usepackage{authblk}
\author[1]{Felix Reichmann*}
\author[1]{Alberto Mistroni}
\author[1]{Fabian Fidorra}
\affil[1]{IHP - Leibniz Institute for High Performance Microelectronics, Frankfurt (Oder) 15236, Germany}
\author[1,2]{Giovanni Capellini}
\affil[2]{Dipartimento di Scienze, Università Roma Tre, Roma 00146, Italy}
\author[1]{Yuji Yamamoto}
\author[1]{Marco Lisker}
\author[1]{Marvin H. Zoellner}

\title{Mobility Enhancement in Si/SiGe Quantum Well Enabled by a Buried Si Layer Trapping Oxygen Impurities}

\date{*Email: reichmann@ihp-microelectornics.com}

\begin{document}

\maketitle

\begin{abstract}

Reducing disorder in undoped Si/SiGe field-effect heterostructures remains an important materials challenge for scalable quantum devices, particularly electron spin qubits. Background impurities such as oxygen have been identified as mobility-limiting, yet practical heterostructure-design strategies for suppressing their incorporation remain underexplored, and their influence across different transport regimes is not fully established. Here, we demonstrate a simple route to oxygen reduction and mobility enhancement in Si/SiGe quantum-well heterostructures grown by reduced-pressure chemical vapor deposition (RP-CVD) on 200~mm Si(100) substrates through the introduction of a thin, electrically passive buried Si layer within the lower SiGe barrier. Secondary-ion mass spectrometry shows that the buried Si layer reproducibly reduces the oxygen background in the subsequently grown SiGe by approximately a factor of five, without modifying the active quantum-well region. Density- and temperature-dependent magnetotransport measurements further show that this reduction increases the electron mobility, while leaving the percolation density and density-dependent mobility scaling largely unchanged. Upon cooling to \(0.3~\mathrm{K}\), both high- and low-oxygen devices exhibit similar density-dependent fractional mobility enhancements, indicating that the reduced oxygen background improves momentum relaxation without substantially altering the dominant low-density disorder landscape. These results establish the buried Si layer as a straightforward and process-compatible heterostructure-design element for reducing oxygen incorporation and improving transport in 200~mm CVD-grown Si/SiGe quantum-device materials.

\end{abstract}

\section*{Keywords}

SiGe heterostructures,  Si quantum wells, chemical vapor deposition, disorder, oxygen incorporation, Hall bar FETs,  electron mobility, semiconductor quantum devices

\section{Introduction}

Undoped Si/SiGe heterostructures are a leading materials platform for field-effect-controlled quantum devices, particularly electron spin qubits, because they combine long coherence times, high tunability, and compatibility with mature CMOS processing \cite{SiSpinQubitsReview2025}.
In these devices, a tensile-strained Si quantum well embedded between relaxed SiGe barriers hosts the buried two-dimensional electron system in which quantum dots are electrostatically confined using nanostructured gates \cite{ScappucciReview2021}. Recent advances include multi-qubit operation \cite{Philips6Qubits2022,FernandezSixQubits2026}, coherent electron shuttling \cite{MusterShuttling2025}, two-qubit operation over extended device distances \cite{Matsumoto2026} and increasingly integrated architectures \cite{NeyensIntel2024,LangheinrichShuttling2025,HRLQPU2026}.

Despite this progress, disorder reduction remains a central challenge for scalable Si/SiGe quantum devices. The quality of both the semiconductor heterostructure and the gate stack influences electrostatic uniformity, reproducible quantum-dot formation, charge noise, and few-electron stability \cite{TaiReview2024,CorleyStrainQBit2023,EspostiMagneto2024,George12Qubits2025, HuckemannQDs2025}. Relevant disorder sources include remote charges near the dielectric interface, strain fluctuations, interface roughness, alloy disorder, and residual background impurities \cite{LarocheQWDepth2015,CorleyStrainShuttler2023,HuangDisorder2023,WuetzNoiseThinQW2023,Mistroni2DEG2024}. Density-dependent transport measurements in Hall-bar field-effect transistors (HB-FETs) provide a rapid means of investigating these contributions before, or alongside, the fabrication of more complex quantum-dot devices \cite{NeyensIntel2024,HRLQPU2026,WuetzMagneto2020,ReichmannHBFET2024}. By tuning the carrier density, such measurements probe mobility, percolation behavior, and quantum-scattering metrics across transport regimes with different sensitivities to the underlying disorder landscape \cite{Esposti2DEG2022,Mistroni2DEG2025}.

This distinction is particularly important for residual background impurities, whose transport signatures can be difficult to isolate from other disorder contributions. Oxygen is known to participate in electrically active defect processes in bulk silicon \cite{ChroneosOxyRev2015}, while in Si and SiGe epitaxy elevated oxygen incorporation has been associated with impaired crystal quality and oxygen-induced defect formation \cite{PivacOxyRev1991,KruegerOxyEpi1994}. This concern extends to undoped group-IV quantum-well heterostructures, where oxygen has been regarded as a potentially detrimental contaminant and its reduction has accompanied the optimization of both Si/SiGe and Ge/SiGe materials platforms \cite{NeyensIntel2024,SammakGeQW2019}. In Ge/SiGe heterostructures, precursor purification reduced the oxygen concentration and increased carrier mobility, while density-dependent transport modeling linked this improvement to a substantial decrease in the fitted background charge density and associated background-impurity scattering \cite{LuOxygenGeQW2026}. In Si/SiGe heterostructures, elevated oxygen concentrations near the quantum well have likewise been correlated with reduced electron mobility \cite{MiScattering2015}. However, the compared Si/SiGe structures also differed in relevant layer dimensions, making it difficult to isolate the transport impact of oxygen from concurrent structural variations. It therefore remains unclear whether oxygen reduction primarily affects the low-density onset of conduction, the mobility at higher carrier density, or the crossover between these regimes. Moreover, because the growth temperature sequence \cite{BedellOxygen2020} and vertical stack design can influence where residual oxygen is incorporated, heterostructure engineering may provide a practical route to suppress oxygen near the active channel without requiring additional reactor- or precursor-purification measures.

Here, we demonstrate such a heterostructure-design approach in 200~mm reduced-pressure chemical vapor deposition (RP-CVD) Si/SiGe quantum-well heterostructures through the introduction of a thin, electrically passive buried Si layer within the lower SiGe barrier. We compare heterostructures with different barrier and quantum-well growth-temperature sequences and show that the buried Si layer reproducibly suppresses the oxygen background in the subsequently grown SiGe without modifying the active quantum-well region. Secondary-ion mass spectrometry (SIMS) measurements establish the resulting changes in the vertical oxygen distribution, while temperature-dependent transport measurements determine how oxygen reduction affects mobility, percolation behavior, and density-dependent transport scaling.

\section{Experimental Section}

To investigate how the Si/SiGe growth-temperature sequence and vertical heterostructure design can be used to locally modify oxygen incorporation, we designed four Si/SiGe heterostructures that vary the Si/SiGe growth-temperature sequence and the presence of an additional buried Si trapping layer. All heterostructures were grown on \(200~\mathrm{mm}\) p-type Si(100) substrates by reduced-pressure chemical vapor deposition (RP-CVD) using silane and germane precursors \cite{ReichmannHBFET2024,Mistroni2DEG2025}. Following a pre-epitaxy wafer clean, a \(4~\mu\mathrm{m}\)-thick step-graded Si\(_{1-x}\)Ge\(_x\) virtual substrate was grown to a nominal Ge concentration of \(x=0.33\pm0.01\), followed by a \(2.7~\mu\mathrm{m}\)-thick Si\(_{0.67}\)Ge\(_{0.33}\) buffer. Chemical-mechanical polishing was applied to reduce the surface roughness associated with cross-hatch formation, followed by a second pre-epitaxy clean prior to growth of the active heterostructure.

Figure~\ref{Figure1}(a) summarizes the relevant layer and growth-temperature sequences of the four investigated designs. Structures without the additional Si trapping layer contain a continuous \(100~\mathrm{nm}\)-thick lower SiGe barrier, whereas structures with the trapping layer contain a \(6~\mathrm{nm}\)-thick Si layer inserted between two \(50~\mathrm{nm}\)-thick lower SiGe barrier sections. The \(8~\mathrm{nm}\)-thick strained-Si QW is buried beneath a \(37~\mathrm{nm}\)-thick upper SiGe barrier. The SiGe barriers were grown at \(600~^\circ\mathrm{C}\), while the Si QW was grown at either \(600~^\circ\mathrm{C}\)  or \(700~^\circ\mathrm{C}\) \cite{Mistroni2DEG2024,Mistroni2DEG2025}. Where present, the Si trapping layer and the Si QW were grown at the same temperature. The four configurations are therefore denoted as ``QW \(700~^\circ\mathrm{C}\)'', ``QW + Trap \(700~^\circ\mathrm{C}\)'', ``QW \(600~^\circ\mathrm{C}\)'', and ``QW + Trap \(600~^\circ\mathrm{C}\)''. All heterostructures were terminated by a \(3~\mathrm{nm}\)-thick epitaxial Si sacrificial cap grown at \(700~^\circ\mathrm{C}\).

Hall-bar field-effect transistors (HB-FETs) were fabricated on the `QW \(600~^\circ\mathrm{C}\)'', and ``QW + Trap \(600~^\circ\mathrm{C}\)'' heterostructure designs using CMOS-compatible processing \cite{ReichmannHBFET2024,FidorraOhmics2026}. Ohmic contacts were formed by selective phosphorus ion implantation with a dose of \(4.5\times10^{15}~\mathrm{cm^{-2}}\) at an implantation energy of \(20~\mathrm{keV}\), followed by a \(1~\mathrm{min}\) activation anneal at \(700~^\circ\mathrm{C}\). A \(10~\mathrm{nm}\)-thick high-density-plasma (HDP) SiO\(_2\) layer deposited at \(300~^\circ\mathrm{C}\) served as the gate dielectric, followed by a \(30~\mathrm{nm}\)-thick TiN top gate deposited by physical vapor deposition. The HB-FETs were structured by reactive ion etching and had a channel width of \(20~\mu\mathrm{m}\) and a voltage-probe spacing of \(300~\mu\mathrm{m}\).

Secondary ion mass spectrometry (SIMS) and magnetotransport measurements were used for compositional and electrical characterization, respectively. Additional details covering SIMS analysis, transport measurements, and data evaluation are provided in the Supporting Information.

\section{Results and Discussion}

\subsection{Heterostructure-Design Control of Oxygen Incorporation}

\begin{figure*}[!t]
    \centering
    \includegraphics[width=0.9\textwidth]{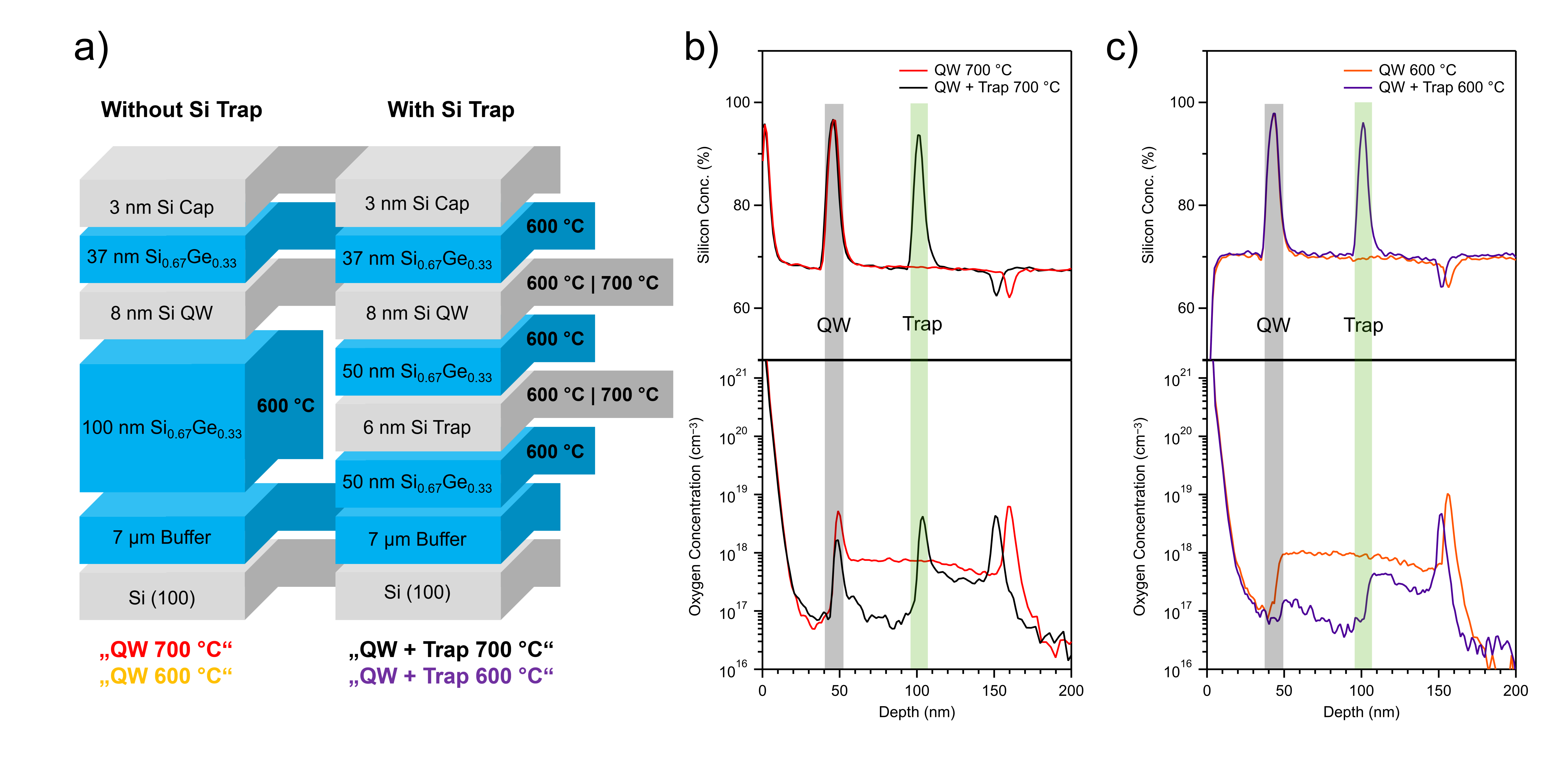}
    \caption{
    Schematic layer sequences and secondary-ion mass spectrometry analysis of oxygen incorporation in Si/SiGe heterostructures with different Si quantum-well (QW) growth temperatures and with or without an inserted Si trap layer. (a) Schematic layer sequences of the heterostructures without and with a buried \(6~\mathrm{nm}\) Si trap layer. The labels below the schematics denote the four investigated growth configurations and correspond to the color coding used in panels (b) and (c). In heterostructures with Si trap, the QW and trap were grown at the same temperature, either at \(600~^\circ\mathrm{C}\) or at \(700~^\circ\mathrm{C}\). (b) Silicon and oxygen depth profiles for structures with a \(700~^\circ\mathrm{C}\)-grown Si QW, comparing samples without and with the additional Si trap layer, shown as red and black lines, respectively. (c) Silicon and oxygen depth profiles for structures with a \(600~^\circ\mathrm{C}\)-grown Si QW, comparing samples without and with the additional Si trap layer, shown as orange and purple lines, respectively.
    }
    \label{Figure1}
\end{figure*}
To determine how the Si/SiGe growth-temperature sequence and the insertion of a buried Si trapping layer affect oxygen incorporation, SIMS depth profiles were acquired for the four heterostructure designs introduced in Figure~\ref{Figure1}(a). The structures are denoted according to the QW growth temperature and the presence of the trapping layer as ``QW \(700~^\circ\mathrm{C}\)'', ``QW + Trap \(700~^\circ\mathrm{C}\)'', ``QW \(600~^\circ\mathrm{C}\)'', and ``QW + Trap \(600~^\circ\mathrm{C}\)''.

We first examine the ``QW \(700~^\circ\mathrm{C}\)'' heterostructure, in which the Si QW is grown at \(700~^\circ\mathrm{C}\) between SiGe barriers grown at \(600~^\circ\mathrm{C}\) \cite{Mistroni2DEG2024,ReichmannHBFET2024}. Figure~\ref{Figure1}(b) shows the corresponding Si and oxygen depth profiles as red lines. The oxygen profile exhibits an elevated background concentration of approximately \(7\times10^{17}~\mathrm{cm^{-3}}\) in the lower SiGe barrier, followed by a pronounced peak reaching approximately \(5\times10^{18}~\mathrm{cm^{-3}}\) near the QW. In the subsequently grown upper SiGe barrier, the oxygen concentration decreases sharply to below \(1\times10^{17}~\mathrm{cm^{-3}}\). This pronounced change across the QW growth step demonstrates that the oxygen incorporation in subsequently grown SiGe is strongly affected by the preceding Si/SiGe growth sequence.

To test whether this behavior can be exploited through the heterostructure design, we introduce an otherwise nominally identical heterostructure containing an additional Si trap layer, also grown at \(700~^\circ\mathrm{C}\) (``QW + Trap \(700~^\circ\mathrm{C}\)'', black profiles). Figure~\ref{Figure1}(b) shows that the inserted Si growth step reduces the oxygen concentration in the subsequently grown part of the lower SiGe barrier from approximately \(4\times10^{17}~\mathrm{cm^{-3}}\) before the trap to \(8\times10^{16}~\mathrm{cm^{-3}}\) after the trap, corresponding to an approximately fivefold reduction. At the same time, pronounced oxygen peaks occur again near both the Si trap layer and the QW, despite the reduced oxygen background between the two Si layers. The persistence of these local oxygen peaks motivates separating the effect of the Si growth step itself from that of the accompanying growth-temperature transition. We therefore repeat the comparison using a matched Si/SiGe growth temperature.

Figure~\ref{Figure1}(c) compares the corresponding ``QW \(600~^\circ\mathrm{C}\)'' (orange profiles) and ``QW + Trap \(600~^\circ\mathrm{C}\)'' (purple profiles) heterostructures, in which the Si QW and, where present, the Si trap layer are grown at \(600~^\circ\mathrm{C}\), matching the growth temperature of the surrounding SiGe barriers \cite{Mistroni2DEG2025}. This matched growth-temperature sequence suppresses the pronounced oxygen peaks observed for the \(700~^\circ\mathrm{C}\) Si growth steps. Within the ``QW + Trap \(600~^\circ\mathrm{C}\)'' heterostructure, the Si trap layer reduces the oxygen concentration in the subsequently grown SiGe from approximately \(4\times10^{17}~\mathrm{cm^{-3}}\) before the trap to \(7\times10^{16}~\mathrm{cm^{-3}}\) after the trap, corresponding to an approximately sixfold reduction. Notably, this reduction is comparable to the approximately fivefold decrease observed for the ``QW + Trap \(700~^\circ\mathrm{C}\)'' heterostructure, indicating that the reduction of the oxygen background following the additional Si growth step persists across both investigated growth-temperature sequences.

The combined SIMS results therefore identify two complementary heterostructure-design elements for reducing oxygen incorporation near the active QW: matching the SiGe-barrier and Si-QW growth temperatures to suppress local oxygen peaks, and introducing an earlier Si growth step within the lower barrier to reduce the oxygen background in subsequently grown layers. For the two \(600~^\circ\mathrm{C}\) heterostructures selected for transport characterization, the oxygen concentration near the QW is approximately \(1\times10^{18}~\mathrm{cm^{-3}}\) for the ``QW \(600~^\circ\mathrm{C}\)'' structure and \(1.3\times10^{17}~\mathrm{cm^{-3}}\) for the ``QW + Trap \(600~^\circ\mathrm{C}\)'' structure. These two structures therefore provide a direct comparison between otherwise similar high- and low-oxygen heterostructure designs. Based on this SIMS contrast, the following transport analysis refers to the ``QW \(600~^\circ\mathrm{C}\)'' structure as the ``high-oxygen'' design and the ``QW + Trap \(600~^\circ\mathrm{C}\)'' structure as the ``low-oxygen'' design.

\subsection{Transport Signatures of Oxygen Reduction at \(1.5~\mathrm{K}\)}

We next investigate the impact of oxygen reduction on carrier-density-dependent transport in Si/SiGe HB-FETs by comparing devices from the high- and low-oxygen heterostructures based on density-dependent mobility, percolation density, and density-dependent mobility scaling.

\begin{figure*}
    \centering
    \includegraphics[width=0.8\textwidth]{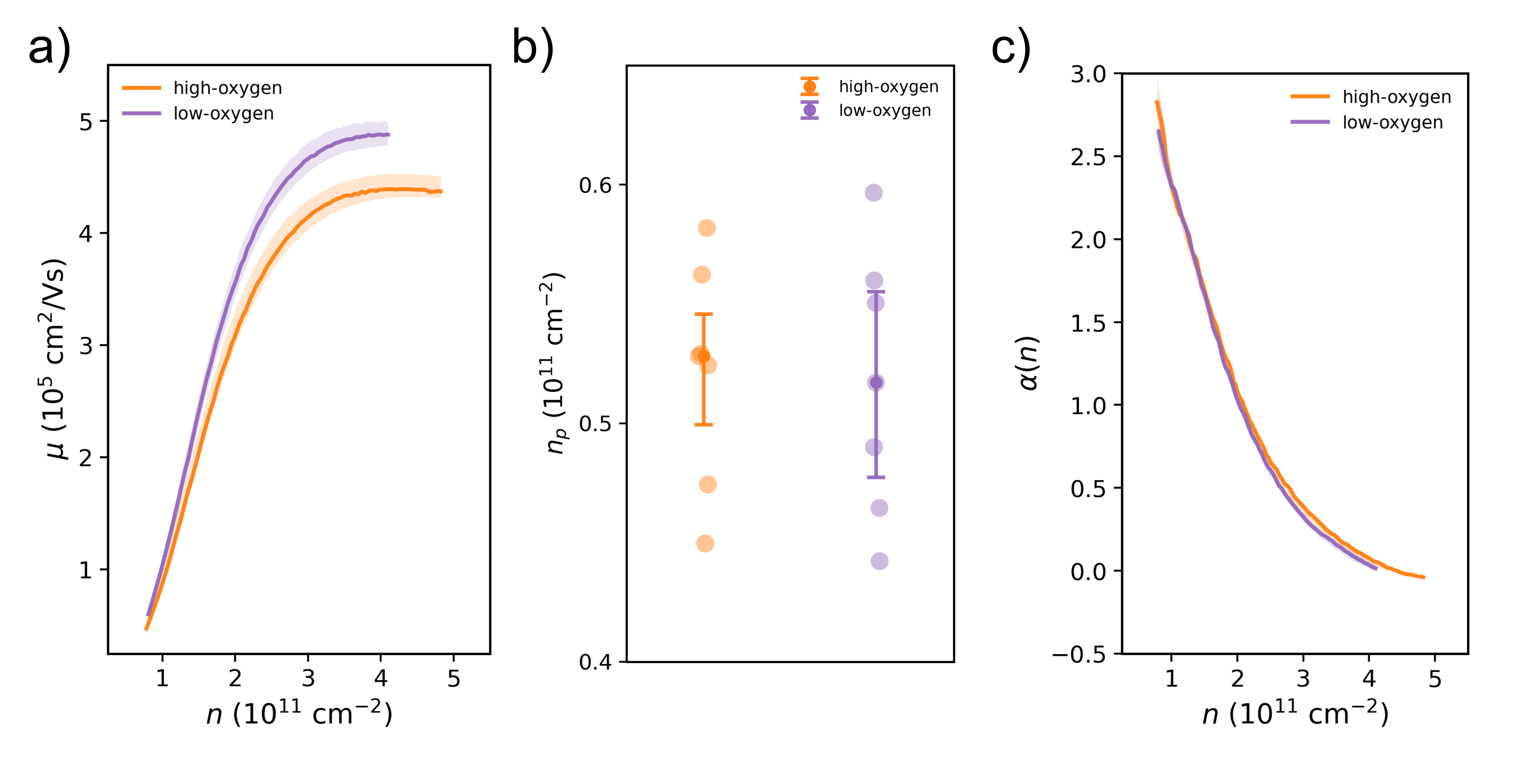}
    \caption{
    Density-dependent transport comparison of high-oxygen (orange) and low-oxygen (purple) Si/SiGe HB-FETs at \(1.5~\mathrm{K}\). Seven devices have been investigated from each group. (a) Median mobility as a function of carrier density. Shaded regions indicate the 25th--75th percentile range within each group. (b) Percolation density \(n_\mathrm{p}\) extracted from the low-density conductivity regime. Individual devices are shown as data points, with the box plot indicating the device-to-device distribution for each group. (c) Median density-dependent mobility exponent \(\alpha(n)\), obtained by first extracting \(\alpha(n)\) from each raw mobility curve using the logarithmic derivative and then taking the median across devices.
    }
    \label{Figure2}
\end{figure*}

Figure~\ref{Figure2}(a) shows the density-dependent mobility of the high-oxygen (orange) and low-oxygen (purple) device groups at \(1.5~\mathrm{K}\), with seven devices characterized for each heterostructure. The curves show the group median, while the shaded regions indicate the 25th--75th percentile spread. Both heterostructures exhibit the same qualitative density dependence. Starting from the low-density conduction regime, the mobility increases strongly with increasing carrier density as enhanced screening progressively reduces the influence of long-range Coulomb disorder \cite{HuangDisorder2023}. At higher electron density, the mobility increase becomes weaker and approaches a maximum around \(n\approx4\times10^{11}~\mathrm{cm^{-2}}\). In this regime, scattering from disorder sources located within or close to the QW, including residual background impurities, interface roughness, and alloy disorder, becomes increasingly relevant \cite{HuangDisorder2023}.

Despite this similar qualitative behavior, a systematic difference develops between the two heterostructure groups as the carrier density increases. Close to the conduction onset, the median mobilities remain comparable, whereas at higher density the low-oxygen devices consistently exhibit higher mobility. The low-oxygen device group reaches a maximum median mobility of approximately \(4.8\times10^{5}~\mathrm{cm^{2}\,V^{-1}\,s^{-1}}\), compared with approximately \(4.3\times10^{5}~\mathrm{cm^{2}\,V^{-1}\,s^{-1}}\) for the high-oxygen device group.

The fact that the mobility enhancement emerges predominantly at higher carrier density, while the two device groups remain comparable close to the conduction onset, suggests that oxygen reduction may have only a limited effect on the disorder governing the low-density transport regime. To examine this more directly, we extract the percolation density \(n_\mathrm{p}\) from the density-dependent longitudinal conductivity \(\sigma_{xx}(n)\) according to the relation
\[
\sigma_{xx} \propto (n-n_\mathrm{p})^{1.31}
\]
\cite{TraceyPercolation2009,Mistroni2DEG2025}. The percolation density is commonly used as a disorder metric relevant for quantum-device operation because it probes the low-density transport regime, where conduction is particularly sensitive to long-range potential fluctuations associated with charged impurities \cite{WuetzMagneto2020,WuetzNoiseThinQW2023}. Figure~\ref{Figure2}(b) shows the distributions of \(n_\mathrm{p}\) for the high-oxygen and low-oxygen device groups. The extracted median percolation densities are nearly identical, with \(5.3\times10^{10}~\mathrm{cm^{-2}}\) for the high-oxygen device group and \(5.2\times10^{10}~\mathrm{cm^{-2}}\) for the low-oxygen device group. Within the fitting accuracy and device-to-device variability, no systematic reduction of \(n_\mathrm{p}\) is therefore observed, indicating that the reduced oxygen concentration near the QW does not significantly affect the low-density conduction onset. This suggests that oxygen reduction does not substantially modify the long-range charged-disorder contribution governing transport in this regime.

However, the mobility improvement observed in Fig.~\ref{Figure2}(a) raises the complementary question of whether oxygen reduction changes the density-dependent mobility scaling or primarily produces an upward shift in the mobility magnitude. We therefore characterize the density dependence using the mobility exponent \(\alpha\), conventionally defined through
\[
\mu \propto n^{\alpha}.
\]
Rather than extracting a single exponent by fitting a broad density range \cite{MonroeScattering1993,LarocheQWDepth2015}, we estimate the logarithmic derivative
\[
\alpha(n)=\frac{d\log(\mu)}{d\log(n)}
\]
from the measured density-dependent mobility, as described in the Methods and shown in Fig.~\ref{Figure2}(c). The median \(\alpha(n)\) curves decrease with increasing carrier density and nearly overlap for the high-oxygen and low-oxygen device groups across the measured density range, indicating that oxygen reduction does not substantially change the density-dependent mobility scaling in these heterostructures.

Together, the \(1.5~\mathrm{K}\) benchmarking results show that reducing the oxygen concentration near the QW increases the absolute mobility, while leaving the percolation threshold and the local density-dependent mobility scaling essentially unchanged. This indicates that the reduced oxygen background improves the mobility without strongly modifying the disorder landscape that controls the onset of conduction and the low-density scaling of \(\mu(n)\). 

However, the absence of a pronounced oxygen-dependent signature in the low-density transport metrics at \(1.5~\mathrm{K}\) does not necessarily imply that the corresponding disorder contributions are indistinguishable at lower temperatures. In Si/SiGe 2DEGs, the density-dependent mobility can remain strongly temperature dependent in the low-density regime, while cooling further into the degenerate regime reduces thermal contributions to the measured transport response \cite{MiScattering2015,VisserMistronimKBarrier2026}. We therefore extend the comparison toward the millikelvin regime to examine whether differences associated with the oxygen background become more apparent at lower temperatures.

\subsection{Impact of Oxygen Reduction on Temperature-Dependent Transport}

\begin{figure*}
    \centering
    \includegraphics[width=0.8\textwidth]{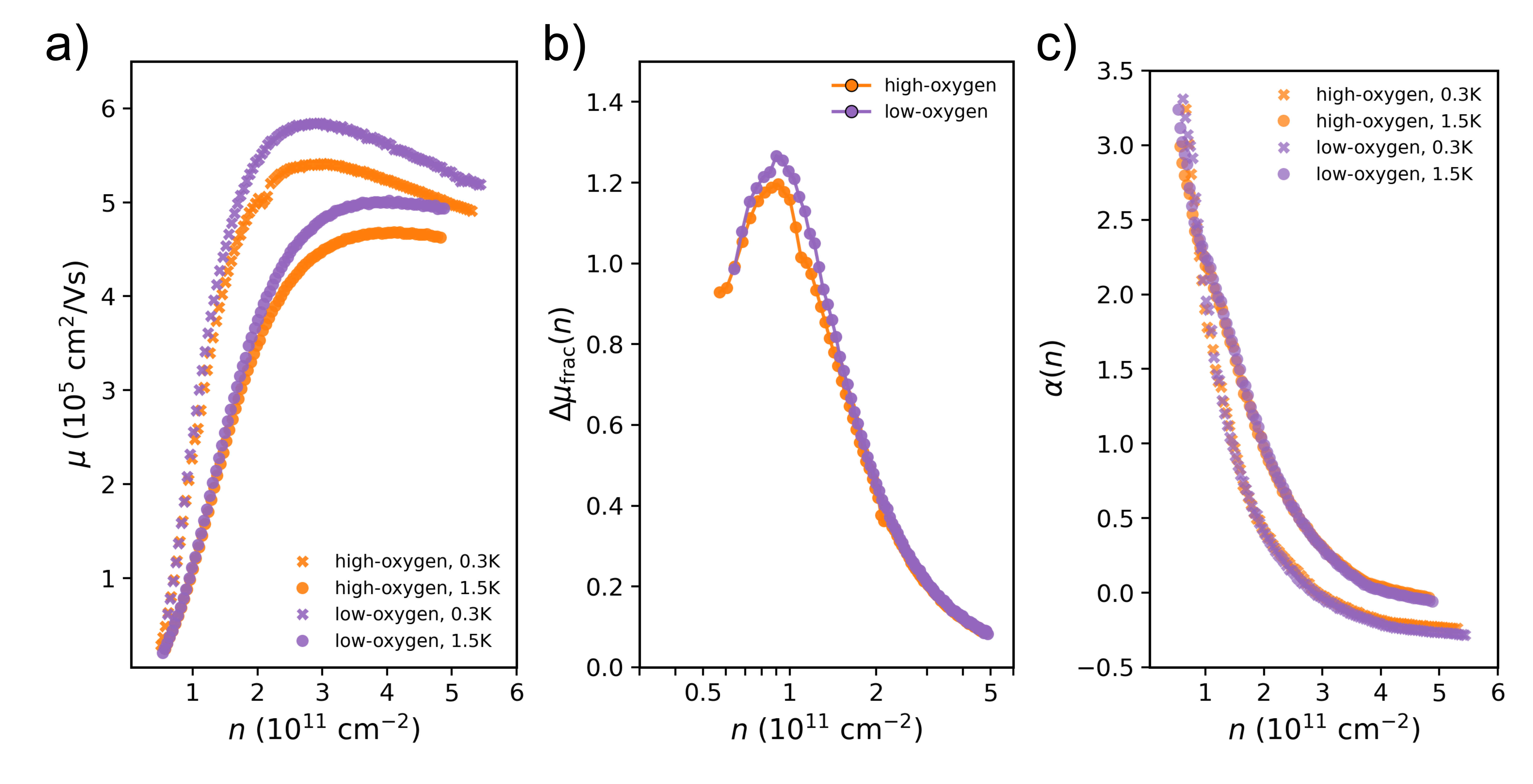}
    \caption{
    Temperature-dependent transport response of representative high-oxygen (orange) and low-oxygen (purple) Si/SiGe HB-FETs upon cooling from \(1.5~\mathrm{K}\) (dots) to \(0.3~\mathrm{K}\) (crosses). (a) Mobility as a function of carrier density. (b) Fractional mobility gain \(\Delta\mu_{\mathrm{frac}}(n)\) upon cooling from \(1.5~\mathrm{K}\) to \(0.3~\mathrm{K}\). (c) Density-dependent mobility exponent \(\alpha(n)\), extracted from the logarithmic derivative of the mobility curves.
    }
    \label{Figure3}
\end{figure*}

To examine the impact of oxygen reduction on temperature-dependent transport, we compare representative high-oxygen and low-oxygen devices at \(1.5~\mathrm{K}\) and \(0.3~\mathrm{K}\). Figure~\ref{Figure3}(a) shows the corresponding density-dependent mobility. At both characterization temperatures, the low-oxygen device exhibits a higher mobility than the high-oxygen device, consistent with the statistical comparison at \(1.5~\mathrm{K}\) in the previous section. However, upon cooling, the mobility increases in both devices over a broad density range and the enhancement appears to vary with carrier density rather than producing a simple density-independent offset. This is also reflected in the shift of the mobility maximum in both devices toward lower density, from approximately \(3.8\times10^{11}~\mathrm{cm^{-2}}\) at \(1.5~\mathrm{K}\) to about \(2.8\times10^{11}~\mathrm{cm^{-2}}\) at \(0.3~\mathrm{K}\). 

To quantify and compare the relative enhancement between the low- and high-oxygen device, we evaluate the fractional mobility gain \(\Delta\mu_{\mathrm{frac}}(n)\) between \(1.5~\mathrm{K}\) and \(0.3~\mathrm{K}\), defined as
\[
\Delta\mu_{\mathrm{frac}}(n)
=
\frac{\mu(0.3~\mathrm{K},n)-\mu(1.5~\mathrm{K},n)}
{\mu(1.5~\mathrm{K},n)} .
\]
The resulting density dependence is shown in Fig.~\ref{Figure3}(b). In both devices, the fractional mobility gain initially increases above the low-density conduction onset, reaches a maximum around \(n=1\times10^{11}~\mathrm{cm^{-2}}\), and subsequently decreases toward higher carrier density. In general, this non-monotonic behavior shows that the mobility enhancement upon cooling is strongly density dependent and is most pronounced in the intermediate-density regime above conduction onset. Such a density-dependent temperature response is consistent with a screening-based picture of low-density 2D transport governed by long-range charged disorder, where the mobility remains temperature sensitive near the onset of conduction and the cooling-induced enhancement weakens as carrier density and screening increase \cite{DasSarma2DEG2015,VisserMistronimKBarrier2026}. However, the quantitative comparison of the two devices through the fractional mobility gain also demonstrates, that oxygen reduction does not produce an additional relative mobility enhancement upon cooling.

We next return to the absolute mobility curves in Fig.~\ref{Figure3}(a) to examine whether oxygen reduction modifies the density-dependent scaling of \(\mu(n)\). At both \(1.5~\mathrm{K}\) and \(0.3~\mathrm{K}\), the high-oxygen and low-oxygen mobility curves remain comparatively close up to a carrier density of approximately \(1.5\times10^{11}~\mathrm{cm^{-2}}\). Above this density, the mobility of the low-oxygen device becomes noticeably larger. This apparent density-dependent separation raises the question of whether oxygen reduction modifies the power-law scaling of the mobility, \(\mu \sim n^\alpha\), rather than merely increasing the absolute mobility.

To test this, we extract the mobility exponent according to \(\alpha(n)=d\log(\mu)/d\log(n)\), as in the previous section and shown in Fig.~\ref{Figure3}(c). While cooling to \(0.3~\mathrm{K}\)  modifies the \(\alpha(n)\) dependence with respect to  \(1.5~\mathrm{K}\), the high-oxygen and low-oxygen devices exhibit very similar \(\alpha(n)\) traces at each respective measurement temperature. Thus, although the absolute mobility of the low-oxygen device becomes noticeably larger above approximately \(1.5\times10^{11}~\mathrm{cm^{-2}}\) in Fig.~\ref{Figure3}(a) , this separation is not accompanied by a pronounced difference in the density-dependent mobility exponent. The comparison of \(\alpha(n)\) for the two representative devices therefore indicates that the reduced oxygen background does not substantially alter the density-dependent power-law scaling of \(\mu(n)\), despite the increased mobility of the low-oxygen device.

Overall, the temperature-dependent measurements reveal a pronounced mobility enhancement upon cooling into the sub-kelvin regime in both of the representative high- and low-oxygen devices. However, the fractional mobility gain shows no clear dependence on the oxygen background. Likewise, the comparison of \(\alpha(n)\) indicates that oxygen reduction does not substantially modify the density-dependent scaling of \(\mu(n)\), despite the higher absolute mobility of the low-oxygen device. Together with the \(1.5~\mathrm{K}\) benchmarking, these observations indicate that reducing the oxygen background primarily enhances the mobility magnitude, while leaving the characteristic density- and temperature-dependent transport response largely unchanged.

\section{Conclusion}

In this work, we investigated how heterostructure design can be used to reduce oxygen incorporation in \(200~\mathrm{mm}\) RP-CVD-grown Si/SiGe field-effect stacks and how this reduction affects density- and temperature-dependent transport.

SIMS measurements show that the oxygen profile near the active QW is sensitive to the Si/SiGe growth-temperature sequence, consistent with trends anticipated from earlier studies of oxygen incorporation during CVD growth of Si, Ge, and SiGe. More importantly, we identify a direct and reproducible heterostructure-design approach for reducing the oxygen background in subsequently grown SiGe through the insertion of a thin, electrically passive Si layer within the lower SiGe barrier. In the \(600~^\circ\mathrm{C}\) structures, this reduction is achieved while maintaining the same growth temperature throughout the relevant Si and SiGe layers, allowing the effect of the insertion layer to be separated from changes associated with the growth-temperature sequence. The resulting heterostructures therefore provide a controlled platform for examining how a substantially reduced oxygen background near the active channel affects electronic transport.

This configuration enabled a direct transport comparison between two otherwise closely related QW heterostructures. For these structures, the oxygen concentration near the QW is reduced from approximately \(1\times10^{18}\) to \(1.3\times10^{17}~\mathrm{cm^{-3}}\), while the maximum mobility at \(1.5~\mathrm{K}\) increases from approximately \(4.3\times10^{5}\) to \(4.8\times10^{5}~\mathrm{cm^2\,V^{-1}\,s^{-1}}\), corresponding to an improvement of about \(11\%\). In contrast, the percolation threshold and density-dependent mobility scaling show no systematic oxygen-dependent change at \(1.5~\mathrm{K}\). Extending the comparison into the sub-kelvin regime leads to the same qualitative picture: although both representative devices exhibit a pronounced mobility enhancement upon cooling, their fractional mobility gain and density-dependent mobility scaling remain comparable. Thus, the substantial change in oxygen concentration primarily affects the absolute mobility, rather than producing a corresponding change in the characteristic low-density or temperature-dependent transport response.

Overall these results identify and confirm oxygen near the QW as a mobility-limiting background impurity, while indicating that oxygen-related disorder does not dominate the long-range potential fluctuations governing the low-density transport regime. The electrically passive Si insertion layer therefore provides a simple, practical and process-compatible heterostructure-design strategy for reducing the oxygen background and improving the transport quality of  RP-CVD-grown Si/SiGe stacks. Whether this reduction also translates into improved few-electron quantum-device operations remains to be established through direct measurements.

\section*{Acknowledgements}

Parts of this work are founded by SPINS, which is co-funded by the Chips Joint Undertaking, under grant agreement No. 101288553, and the National Funding Authorities of the Participating States: Belgium, Czechia, Finland, France, Germany, Italy, the Netherlands, Portugal, and Spain.

\section*{Author Contributions}

\textbf{F. R.:} Writing – original draft \& review \& editing, Conceptualization, Investigation, Visualization, Formal analysis;
\textbf{A. M.}:  Writing – review \& editing, Investigation;
\textbf{F. F.}: Writing – review \& editing; 
\textbf{G. C.}: Writing – review \& editing;
\textbf{Y. Y.}: Writing – review \& editing, Resources;
\textbf{M. L.}: Writing – review \& editing, Resources;
\textbf{M. H. Z.}: Writing – review \& editing, Conceptualization, Formal analysis, Funding acquisition, Project administration.

\section*{Data Availability}

The data that support the findings of this study are available from the corresponding author upon request.

\section*{Conflict of Interest}

The authors declare no conflict of interest.

\bibliography{references}

\newpage

\section*{Supporting information}

\setcounter{figure}{0}
\renewcommand{\thefigure}{S\arabic{figure}}

\subsection{Secondary Ion-Mass Spectroscopy}

Secondary ion mass spectrometry (SIMS) depth profiles were acquired using a Physical Electronics ADEPT 1010 dynamic SIMS instrument by the eurofins Evans Analytical Group (EAG) Laboratories. Depth profiling was performed using Cs as the primary ion source, and all monitored species were acquired simultaneously during the measurement. The analyzed secondary-ion signals included \(^{16}\mathrm{O}\), \(^{70}\mathrm{Ge}\), and \(^{30}\mathrm{Si}\), allowing the oxygen concentration to be correlated with the Si/SiGe heterostructure profile. The primary beam spot size was estimated to be around \(50~\mu\mathrm{m}\). Sputter craters were measured using a stylus profilometer and used for depth calibration. Quantification was performed using Si and SiGe reference materials, with the SiGe reference calibrated by Rutherford backscattering spectrometry. The SIMS data were evaluated using PCOR-SIMS, which adjusts the relative sensitivity factors and sputter rate according to the local Ge concentration in the SiGe matrix.

\subsection{Magnetotransport Characterization}

The \(1.5~\mathrm{K}\) benchmarking measurements were performed in an Oxford Instruments Teslatron-PT cryostat, while the temperature-dependent measurements between \(1.5~\mathrm{K}\) and \(0.3~\mathrm{K}\) were carried out in a Kiutra L-Type Rapid cryostat. After reaching the target temperature, the two-dimensional electron gas in the Si quantum well was accumulated by applying a DC gate voltage \(V_{\mathrm{G}}\) above the threshold voltage \(V_{\mathrm{thr}}\). The current was limited to \(50~\mathrm{nA}\) using a \(1~\mathrm{M}\Omega\) series resistor.

Hall-effect measurements were performed by sweeping the perpendicular magnetic field \(B_{\perp}\) from \(-5~\mathrm{mT}\) to \(305~\mathrm{mT}\). During each field sweep, the longitudinal voltage \(V_{xx}\), transverse voltage \(V_{xy}\), and channel current were recorded using a low-frequency four-probe lock-in technique at \(17~\mathrm{Hz}\) with a Specs Nanonis Tramea system. The carrier density \(n\) was extracted from the Hall slope according to
\[
\rho_{xy} = \frac{B_{\perp}}{en},
\]
where \(e\) is the elementary charge. The low-field mobility was calculated as
\[
\mu = \frac{1}{e n \rho_{xx}(0)},
\]
where \(n\) is the carrier density extracted from the low-field Hall slope and \(\rho_{xx}(0)\) is the longitudinal sheet resistivity evaluated near \(B_{\perp}=0\).  The longitudinal conductivity used for the percolation fit was calculated from the same zero-field longitudinal resistivity according to
\[
\sigma_{xx}(0) = \frac{1}{\rho_{xx}(0)}.
\]
The resulting \(\sigma_{xx}(0)\) was converted to units of \(e^2/h\) and used to extract the percolation density \(n_p\), as described in the main text.

To extract the density-dependent mobility scaling without relying on a point-to-point numerical derivative, \(\alpha(n)\) was estimated by a sliding-window linear least-squares fit of \(\log_{10}(\mu)\) versus \(\log_{10}(n)\). For each density point, the fit was performed within a symmetric window in \(\log_{10}(n)\) with a total width of \(0.22\) decades and a minimum of five data points. Near the boundaries of the measured density range, the five nearest points were used when fewer than five points fell within the symmetric window. The extracted local slope was assigned to the corresponding density. For grouped comparisons, the individual \(\alpha(n)\) traces were interpolated onto a common density grid within the shared density range, and the median and interquartile range were calculated at each density point. The corresponding individual \(\alpha(n)\) traces are provided in the Supplementary Information.

To obtain the fractional mobility gain \(\Delta\mu_{\mathrm{frac}}(n)\), the datasets closest to the target temperatures of \(0.3~\mathrm{K}\) and \(1.5~\mathrm{K}\) were selected for each sample. Because the measured density points generally differ between the two temperature sweeps, the mobility values were evaluated on a common density axis using linear interpolation without extrapolation. The measured density points of the \(1.5~\mathrm{K}\) dataset, corresponding to the denominator in \(\Delta\mu_{\mathrm{frac}}(n)\), were used as the evaluation densities \(n_{\mathrm{eval}}\). Only points within the overlapping density range of the two datasets were retained, such that \(\mu(0.3~\mathrm{K},n_{\mathrm{eval}})\) and \(\mu(1.5~\mathrm{K},n_{\mathrm{eval}})\) were compared at identical carrier densities.

\subsection{Carrier Density Calibration}

\begin{figure}
    \centering
    \includegraphics[width=0.65\linewidth]{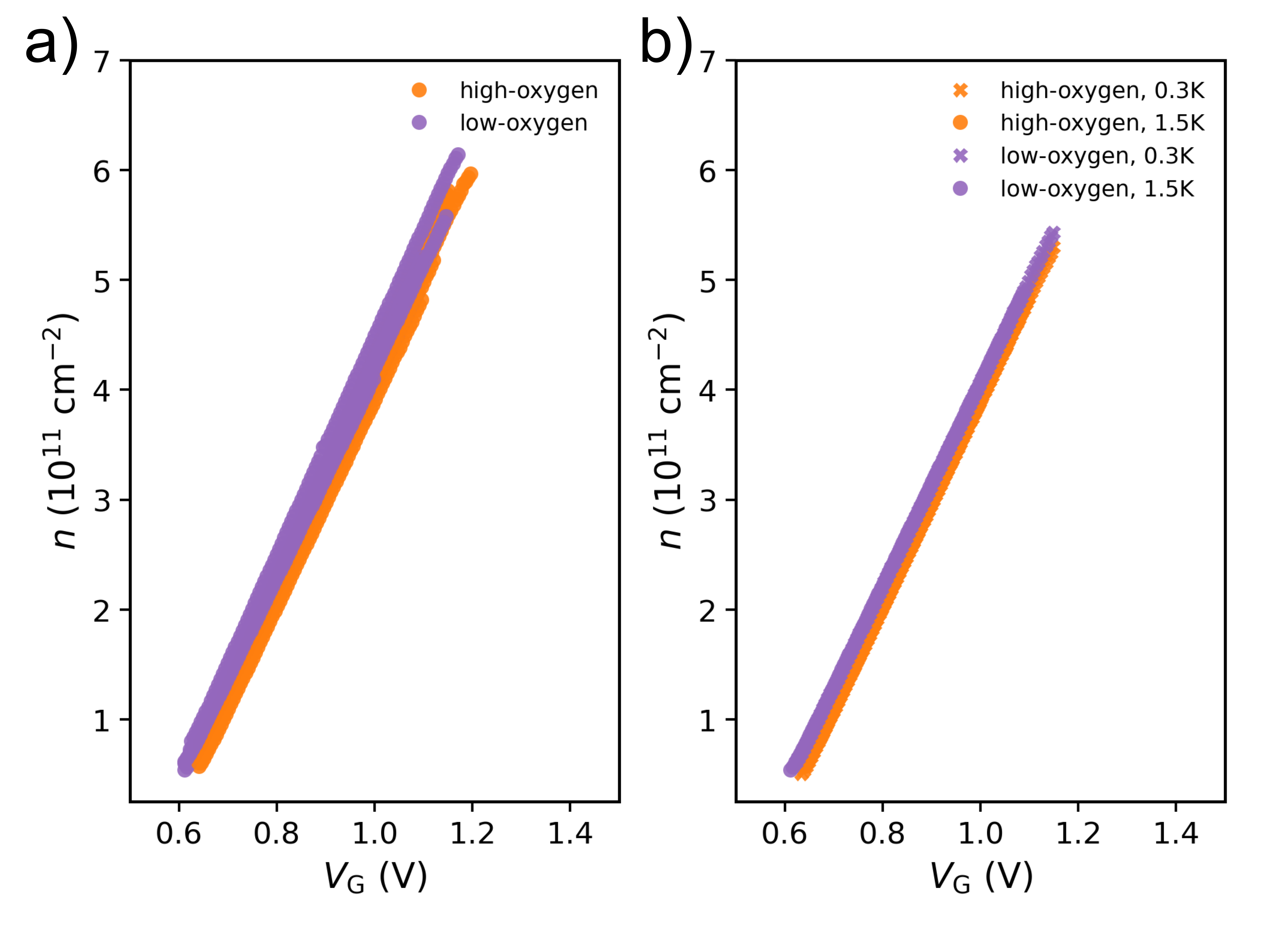}
    \caption{
    Carrier density as a function of gate voltage for devices with high-oxygen (orange) and low-oxygen concentration (purple), displaying the linear range used for the carrier density calibration in the main text. (a) Shows devices included in the 1.5~K benchmark comparison, while (b) shows plots from two selected the devices measured between 1.5~K and 0.3~K.
    }
    \label{SI-Figure1}
\end{figure}

In this section, we show the linear \(n(V_\mathrm{G})\) regime used as the carrier-density range for the transport metrics discussed in the main text. Figure~\ref{SI-Figure1}(a) shows the device-resolved \(n(V_G)\) traces for all devices included in the \(1.5~\mathrm{K}\) benchmark. The high-oxygen and low-oxygen devices exhibit overlapping, approximately linear density responses over the relevant gate-voltage range, indicating comparable electrostatic gate coupling within the expected device-to-device and wafer-to-wafer variation of this platform. Figure~\ref{SI-Figure1}(b) shows the corresponding \(n(V_\mathrm{G})\)traces for the devices used in the temperature-dependent comparison. The density response remains nearly unchanged upon cooling from \(1.5~\mathrm{K}\) to \(0.3~\mathrm{K}\), confirming that the observed temperature-dependent and/or oxygen reduction induced mobility enhancement is not caused by a change in density calibration or gate coupling.

\subsection{Raw 1.5 K Benchmark Data}

\begin{figure}
    \centering
    \includegraphics[width=0.7\linewidth]{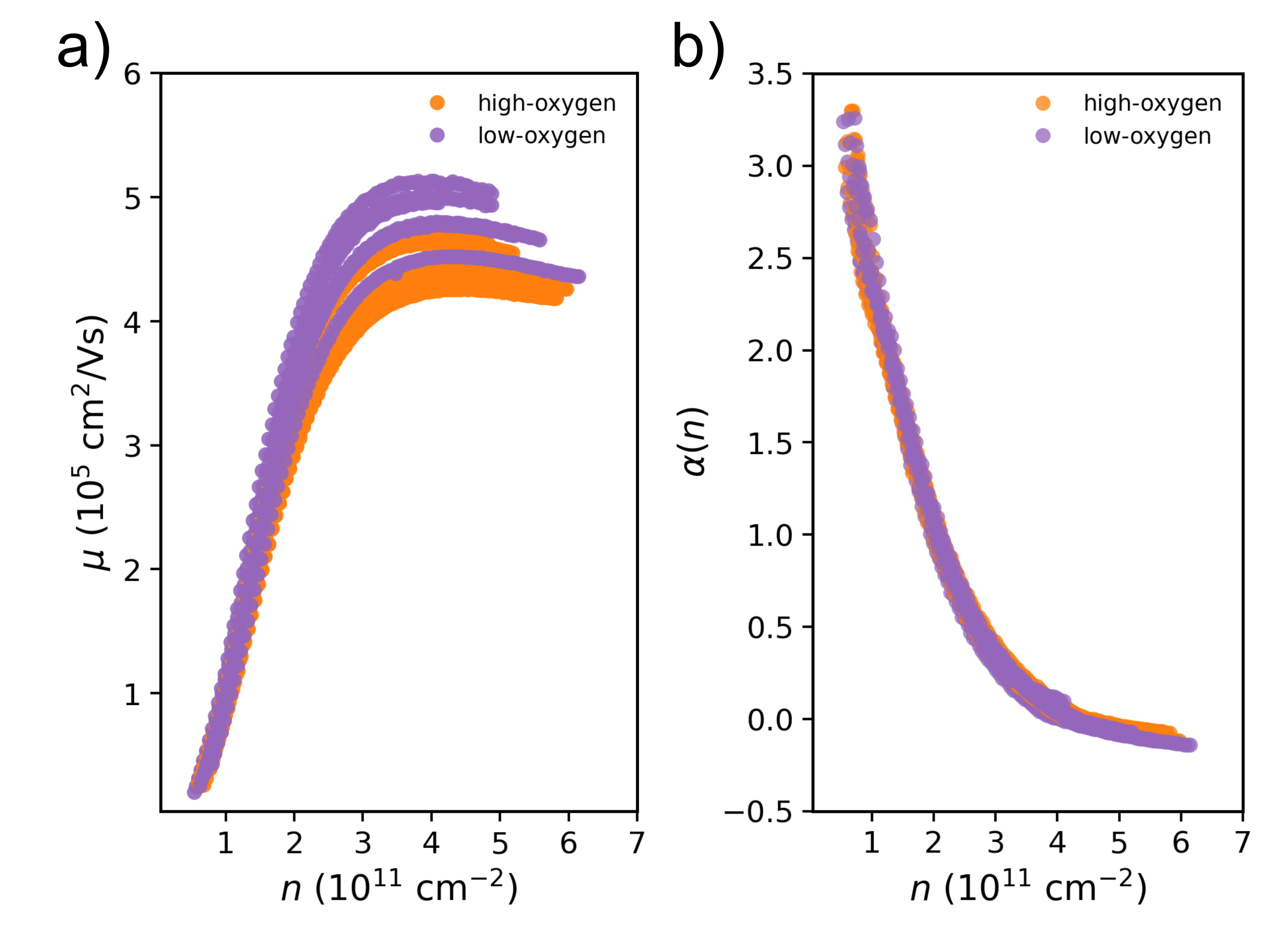}
    \caption{
    Raw mobility and mobility-exponent data sets, for high-oxygen (orange) and low-oxygen (purple) devices , used  to calculate the mean values of the \(1.5~\mathrm{K}\) benchmark data in the main text.
    (a) Raw mobility \(\mu(n)\) traces for high-oxygen and low-oxygen devices plotted as a function of Hall density. (b) Corresponding density-dependent mobility exponent \(\alpha(n)\).
    }
    \label{SI-Figure2}
\end{figure}

Here, we show the device-resolved mobility and local mobility-exponent traces obtained from the \(1.5~\mathrm{K}\) benchmark data. Figure~\ref{SI-Figure2}(a) shows the raw \(\mu(n)\) traces for the high-oxygen and low-oxygen devices, plotted together to illustrate the full device-to-device spread. While the low-oxygen devices tend to exhibit higher mobility at elevated carrier density, both groups follow the same overall density dependence and remain comparable at lower density. Figure~\ref{SI-Figure2}(b) shows the density-dependent mobility exponent \(\alpha(n)=d\log(\mu)/d\log(n)\) from the logarithmic derivative of each mobility curve shown Figure~\ref{SI-Figure2}(a). The high-oxygen and low-oxygen devices exhibit strongly overlapping \(\alpha(n)\) characteristics over the analyzed density range, indicating that the oxygen reduction does not systematically modify the mobility-scaling behavior. This supports the interpretation that the main effect of oxygen reduction is an increase in the absolute mobility, rather than a qualitative change in the underlying density-dependent transport regime.

\subsection{Manual Extraction of the Mobility Exponent}

\begin{figure}
    \centering
    \includegraphics[width=0.7\linewidth]{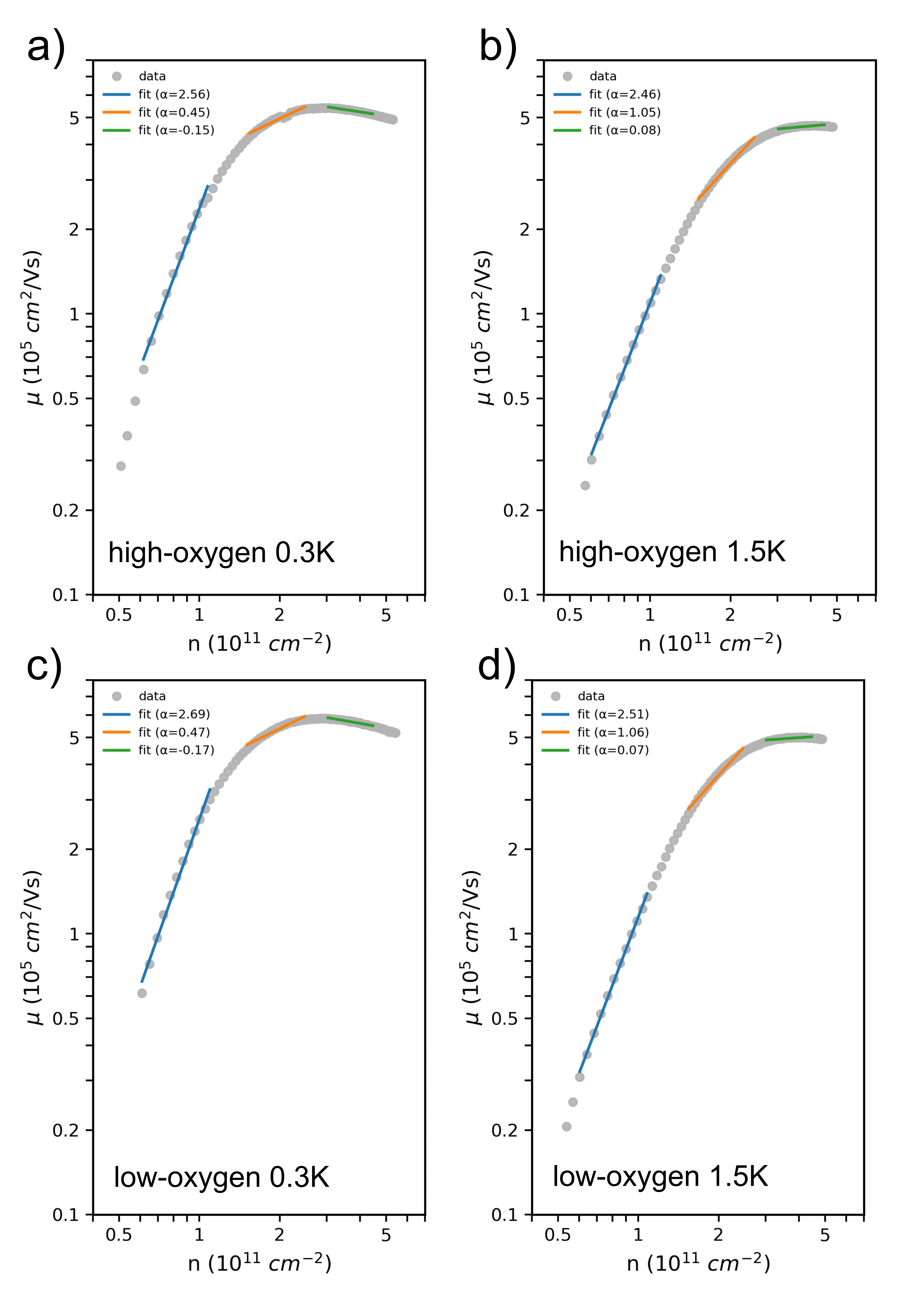}
    \caption{
    Manual extraction of the mobility exponent from double-logarithmic mobility fits.
    The mobility was fitted according to \(\mu \propto n^\alpha\) in three selected density intervals:
    \(0.7{-}1.2\times10^{11}~\mathrm{cm^{-2}}\) (blue),
    \(1.5{-}2.5\times10^{11}~\mathrm{cm^{-2}}\) (orange), and
    \(3.0{-}4.5\times10^{11}~\mathrm{cm^{-2}}\) (green).
    Representative fitting results are shown for a high-oxygen device at
    (a) \(0.3~\mathrm{K}\) and (b) \(1.5~\mathrm{K}\), and for a low-oxygen device at
    (c) \(0.3~\mathrm{K}\) and (d) \(1.5~\mathrm{K}\).
    }
    \label{SI-Figure3}
\end{figure}

A common approach to assess density-dependent mobility scaling is to fit the mobility to a power law, \(\mu \propto n^\alpha\), over selected carrier-density intervals in a double-logarithmic representation. In the main text, we instead introduced the density-dependent mobility exponent defined as \(\alpha(n)=d\log(\mu)/d\log(n)\), obtained from the logarithmic derivative of \(\mu(n)\), to avoid relying on a single subjectively chosen fitting range. To validate this approach and to enable direct comparison with the conventional procedure used in the literature, Figure~\ref{SI-Figure3} shows the result of manual power-law fits performed in three representative carrier-density intervals:
\(0.7{-}1.2\times10^{11}~\mathrm{cm^{-2}}\),
\(1.5{-}2.5\times10^{11}~\mathrm{cm^{-2}}\), and
\(3.0{-}4.5\times10^{11}~\mathrm{cm^{-2}}\).

Figures~\ref{SI-Figure3}(a) and~\ref{SI-Figure3}(b) show the fitting results for a representative high-oxygen device at \(0.3~\mathrm{K}\) and \(1.5~\mathrm{K}\), respectively. Figures~\ref{SI-Figure3}(c) and~\ref{SI-Figure3}(d) show the corresponding analysis for a representative low-oxygen device. For a given temperature, the extracted \(\alpha\) values are comparable between the high-oxygen and low-oxygen devices. This confirms the conclusion obtained from the local logarithmic-derivative analysis in the main text: oxygen reduction increases the absolute mobility, but does not systematically modify the density-dependent mobility scaling.

\subsection{Percolation-Threshold Fitting}

\begin{figure}
    \centering
    \includegraphics[width=0.7\linewidth]{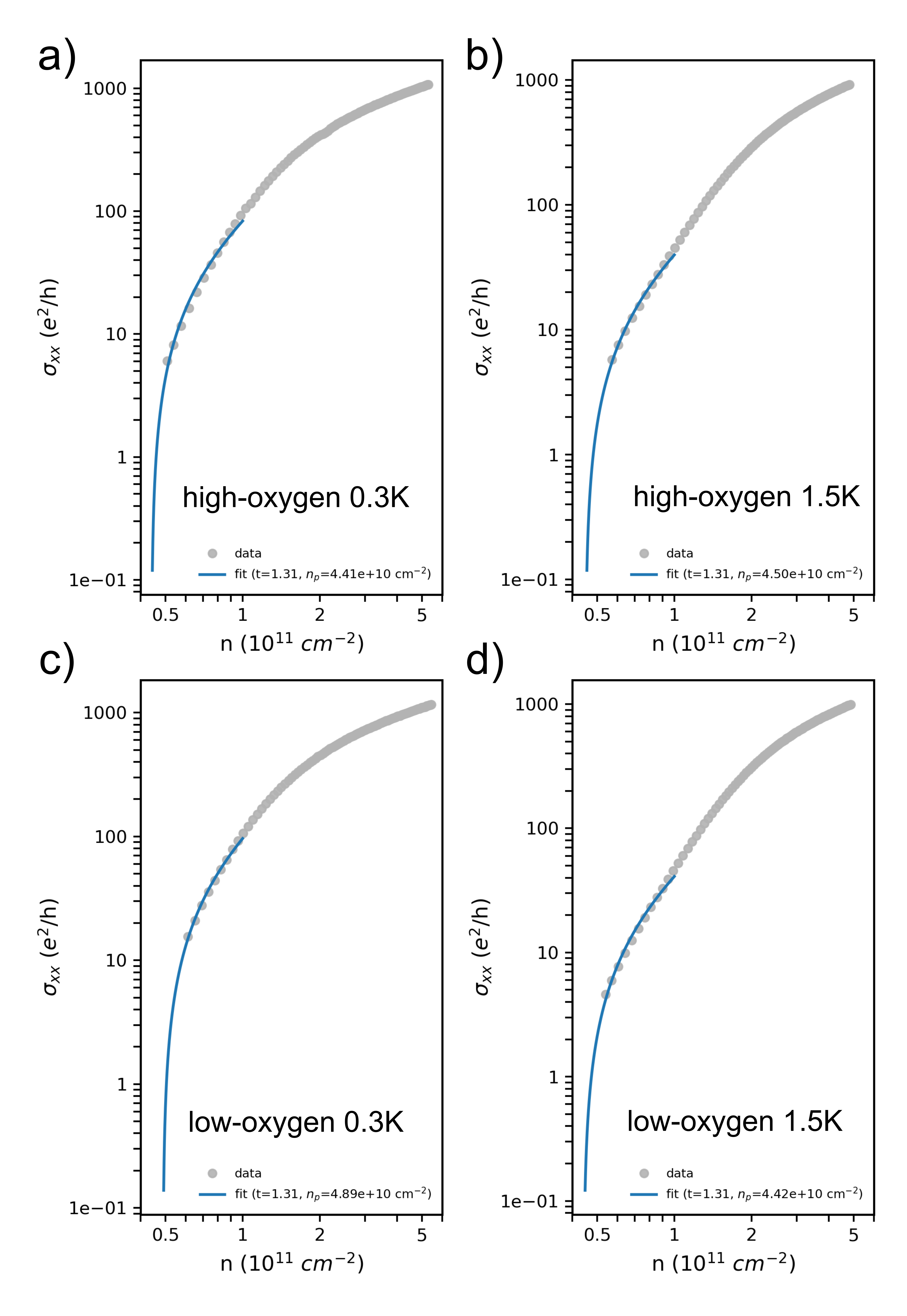}
    \caption{
    Percolation-threshold extraction from the low-density conductivity.
    The conductivity was fitted using a percolation-like power law,
    \(\sigma \propto (n-n_p)^{t}\), over the carrier-density range from
    \(0\) to \(1.2\times10^{11}~\mathrm{cm^{-2}}\).
    Representative fitting results are shown for a high-oxygen device at
    (a) \(0.3~\mathrm{K}\) and (b) \(1.5~\mathrm{K}\), and for a low-oxygen device at
    (c) \(0.3~\mathrm{K}\) and (d) \(1.5~\mathrm{K}\).
    }
    \label{SI-Figure4}
\end{figure}

The percolation threshold \(n_p\) was extracted from the low-density conductivity using a percolation-like power-law fit of the form \(\sigma \propto (n-n_p)^t\) with \(t = 1.31\). Figure~\ref{SI-Figure4} shows representative fits for high-oxygen and low-oxygen devices at \(0.3~\mathrm{K}\) and \(1.5~\mathrm{K}\). For all panels, the fitting range was restricted to carrier densities between \(0\) and \(1.2\times10^{11}~\mathrm{cm^{-2}}\), covering the low-density regime close to the onset of conduction. Figures~\ref{SI-Figure4}(a) and~\ref{SI-Figure4}(b) show the fitting results for a representative high-oxygen device at \(0.3~\mathrm{K}\) and \(1.5~\mathrm{K}\), respectively. Figures~\ref{SI-Figure4}(c) and~\ref{SI-Figure4}(d) show the corresponding analysis for a representative low-oxygen device. Within this fitting procedure, the extracted \(n_p\) values are comparable between the high-oxygen and low-oxygen devices and remain nearly unchanged upon cooling from \(1.5~\mathrm{K}\) to \(0.3~\mathrm{K}\). This confirms that oxygen reduction does not systematically modify the low-density percolative transport regime, supporting the conclusion of the main text.

\end{document}